# An elevation of 0.1 light-seconds for the optical jet base in an accreting Galactic black hole system

P. Gandhi*[,1], M. Bachetti[2], V.S. Dhillon[3,4], R.P. Fender[5], L.K. Hardy[3], F.A. Harrison[6], S.P. Littlefair[3], J. Malzac[7], S. Markoff[8], T.R. Marsh[9], K. Mooley[5], D. Stern[10], J.A. Tomsick[11], D.J. Walton[12], P. Casella[13], F. Vincentelli[14,15,13], D. Altamirano[1], J. Casares[4,16,5], C. Ceccobello[8], P.A. Charles[1,5], C. Ferrigno[17], R.I. Hynes[18], C. Knigge[1], E. Kuulkers[19], M. Pahari[20], F. Rahoui[21], D.M. Russell[22], A.W. Shaw[23]

**Relativistic plasma jets are observed in many accreting black holes. According to theory, coiled magnetic fields close to the black hole accelerate and collimate the plasma, leading to a jet being launched[1–3]. Isolating emission from this acceleration and collimation zone is key to measuring its size and understanding jet formation physics. But this is challenging because emission from the jet base cannot be easily disentangled from other accreting components. Here, we show that rapid optical flux variations from a Galactic black-hole binary are delayed with respect to X-rays radiated from close to the black hole by ~0.1 seconds, and that this delayed signal appears together with a brightening radio jet. The origin of these sub-second optical variations has hitherto been controversial[4–8]. Not only does our work strongly support a jet origin for the optical variations, it also sets a characteristic elevation of ≤$10^3$ Schwarzschild radii for the main inner optical emission zone above the black hole[9], constraining both internal shock[10] and magnetohydrodynamic[11] models. Similarities with blazars[12,13] suggest that jet structure and launching physics could potentially be unified under mass-invariant models. Two of the best-studied jetted black hole binaries show very similar optical lags[8,14,15], so this size scale may be a defining feature of such systems.**

In June 2015, the Galactic X-ray binary V404 Cygni underwent the brightest outburst of an X-ray binary so far this century. We coordinated simultaneous optical observations from the William Herschel Telescope with X-ray observations from the *NuSTAR* space observatory on the morning of June 25. These were high frame-rate optical observations taken by the ULTRACAM instrument, sampling timescales down to 35.94 milliseconds (ms). Both optical and X-ray light curves show variability on a broad range of timescales characteristic of this source[15,16] (Fig. 1). The AMI telescope provided contiguous radio coverage throughout this period. Details of the observations may be found in Methods.

These coordinated observations occurred on June 25, the day preceding the peak of the 2015 outburst. When the optical observations began, the X-ray intensity was two orders of magnitude below peak, and the spectrum was dominated by low-energy X-rays (i.e., it was in a state characterised as being relatively 'soft'). Steady, compact jet activity is not expected in such a state, and consistent with this, the radio spectral index is negative, as is typical of emission from discrete optically-thin ejecta. *NuSTAR* observations were interrupted about 2000 seconds later due to a period of Earth occultation, which separates the two halves (hereafter, 'epochs') of the sequence under consideration.

At some point during this occultation, the source underwent a dramatic and very rapid change in its X-ray spectral state. When *NuSTAR* emerged from Earth occultation, the spectrum was instead found to have pivoted towards high energies, with a larger fraction of X-ray counts above 10 keV than below (i.e. the state was significantly 'harder'). In addition, there is a sharp rise in radio flux as well as spectral index, all of which signal the strengthening of compact radio jet emission in an X-ray hard state. The fractional r.m.s. amplitude characterising the strength of the variations also increases in both optical and X-rays. Together, these facts indicate that we serendipitously caught a rapid state transition, with a sharp divide in its observed properties between the first and second epochs. This was the final transition leading up to the overall outburst peak on the following day (Fig. 2), when the source displayed enhanced gamma-ray emission[17], exceedingly bright X-ray[16] and radio[15] emission, as well as pronounced sub-second optical flaring[15]. Pinpointing any one instant as the moment of the transition is not possible, but we note the sudden appearance of a short-lived rapid optical flaring episode about midway into the occultation (Fig. 1a), which may be regarded as the likely point of transition.

Cross-correlating the light curves reveals the short optical time delay with respect to X-rays (Fig. 1). Whereas no lag is apparent during the first epoch, a significant signal appears in the timing correlation function during the second epoch. The signal is skewed towards positive optical lags, with a weighted time delay $\tau=+0.13\pm0.04$ s of the optical with respect to X-rays. The delay is present in shorter light curve segments, and also in the Fourier domain lags, confirming its robust detection (Supplementary Information). The percentage probability of the signal arising by chance is less than 1 %. Correlations between the optical and X-rays on longer (seconds to minutes) timescales have already been reported[18,19], and are visually apparent for the broad flares in Fig. 1 also. The rapid and regular cadence of our observations, combined with exquisite data quality due to the brightness of the outburst, has allowed us to push time lag detections to the sub-second regime.

A handful of other black hole binaries are known to show significant optical variability, including both positive and negative responses to X-ray variations, on characteristic timescales of order seconds and less[20]. Their origin remains under debate, with possibilities including a magnetically-active corona, an advective flow, a jet or a combination of components[4—8]. Evidence to distinguish between these models has been lacking so far.

We now have this evidence for V404 Cygni. There is a clear change in the timing correlation between the optical and X-rays as a function of radio flux across the state transition. The sub-second delay is only detected during the second epoch, when both radio flux and spectral index rose sharply. This strongly supports a causal link between the jet and the sub-second fluctuations. The delay $\tau$ corresponds to a maximal distance scale of order $c\tau=1.4(\pm0.4)\times10^3$ Schwarzschild radii ($R_S=2GM_{BH}/c^2$, with G being the Gravitational constant, $c$ the speed of light) for a black hole mass $M_{BH}=9$ $M_\odot$ [21]. The expected timescales for X-ray heating of the outer disc or the companion are of order tens of seconds, much longer than the fast delay under question here. Any reprocessing of X-rays in an axisymmetric disc wind known in this system is also expected to occur on similarly long timescales (Supplementary Information). An inner hot advective flow powering the fast optical and X-ray variations is, likewise, incompatible with the data because it predicts the two bands to be anti-correlated[7]. We therefore associate the inferred compact scale with the inner jet, where optical emission is expected to arise as synchrotron radiation. Detection of a lag alone cannot fully constrain the geometry of the optical emission zone in the inner jet. But additional support for this hypothesis comes from the fact that the sub-second flaring is significantly redder (as expected from optically-thin synchrotron) than other emission components[15]. Moreover, a red spectral shape argues against an origin due to standard thermal reprocessing.[15] Finally, a rise in optical polarisation was detected together with rising radio fluxes during the present outburst[22], interpreted as evidence for optical synchrotron radiation.

Spectral analysis of the *NuSTAR* data constrains the X-ray emission region to also be compact. In particular, the brightest flares appear to be extremely hard spectral events confined to ≲ 5 $R_S$ (ref. 17). The emission source for these X-rays can be attributed to the hot electron corona overlying the disc, or the base of the jet, which may be cospatial. Emission during fainter episodes instead originates in a geometrically thick and absorbed flow[23]. Irrespective of the physical mechanism of X-ray emission, there appears to be no doubt that it must occur in a compact region very close to the black hole.

Based upon the above observational facts, we interpret our measurement of $\tau$ as the characteristic propagation delay between the X-ray and the first optical emission zone (the 'optical base') in the inner jet (Fig. 3). This association is significant because it constrains the physical scale over which plasma is accelerated in the inner jet. The physics of feeding, acceleration and collimation of the plasma within this acceleration and collimation zone (ACZ) is thought to be governed by magnetohydrodynamic (MHD) processes[3]. In the blazar BL Lac, which is a jet-dominant supermassive black hole, synchrotron flaring is observed both within the ACZ at elevations of several hundred $R_S$ above the black hole as radiation from accelerated plasma knots is beamed toward the line-of-sight, and also as the plasma crosses a standing shock at an elevation of ~$10^4$ $R_S$ where the ACZ is estimated to terminate[12]. Disruptions leading to shocks in an MHD flow likely require a combination of being beyond the causality surface for fast magnetosonic waves, and some change in external pressure or other perturbative forces. Where exactly this occurs and why is currently unknown and must be determined observationally, but once determined provides a critical constraint on numerical models. With the X-rays arising from very close to the black hole in V404 Cygni, and the origin of the rapid optical flares being inner jet synchrotron, the simplest interpretation is that our measured delay $\tau$ corresponds to the ACZ crossing time of the plasma (in the observed frame of reference), thus placing a limit of ~$10^3$ $R_S$ on its elevation above the black hole. Some of the flares may instead arise from within the ACZ, as in BL Lac. However, our timing cross-correlations here measure a bulk property of the flow of the continuous jet, rather than the location of any single knot, and thus provide a robust limit to the ACZ elevation in V404 Cygni.

Once a steady jet has been launched, its characteristic broadband emission spectrum must also be explained. These spectra are typically flat in flux density units, which requires that adiabatic expansion losses be somehow compensated for. This has motivated internal shock models in which collisions between plasma shells with a velocity shear provide the requisite energy injection mechanism[10,24]. The location of the optical base of the jet corresponds to the zone where such shocks are first produced. If the jet is fed by the accretion flow, a time lag between the X-ray emitting accretion flow and the shock synchrotron radiation is expected, with the magnitude of this lag depending mostly on the characteristic time scale of velocity fluctuations driving the shocks. Lags of order $\tau$~0.1 s may naturally be obtained under the assumption that the power spectrum of the distribution of shock Lorentz factors is similar to that powering the X-ray emission[10]. However, despite the presence of flux variations on all timescales, the power spectra of the X-ray emission of V404 Cygni during the present observations are dominated by slow fluctuations (Supplementary Information), thus predicting a much longer lag. This is strongly constraining in that it rules out the jet Lorentz factor fluctuations being driven by the observed X-ray variability during this state transition, arguing for a revision of current models. We cannot, however, exclude the possibility that complex absorption[23] or reflection[16] effects are conspiring to suppress the fastest X-ray fluctuations from being observed.

Much of the physics of jets is considered to be scale-invariant, with supermassive black hole jets being scaled up counterparts of the stellar-mass transient jets under consideration here[25,26]. State-of-the-art MHD models allow for a wide range of standing shock elevations, depending upon conditions at the jet formation zone (Ceccobello et al. submitted). Our measured elevation of ≲$10^3$ $R_S$ in V404 Cygni is similar to (though perhaps closer to the lower end of) the scales over which flaring occurs in BL Lac, suggestive of mass-scaled similarities in the relevant MHD physics. Whether it is a combination of MHD effects and environmental pressure or some other perturbation that governs this location will need to be explored. Our work provides the boundary conditions enabling such an investigation.

Only two other transient black hole binaries with strong jet optical emission have high-quality timing observations allowing a detailed search for sub-second correlation signals. These are XTE J1118+480[27] and GX 339-4[8], both with strictly simultaneous data in both optical and X-rays at high time resolution. However, there are important differences with respect to our work reported here. XTE J1118+480 was observed in a different state during outburst decline with falling optical, infrared and X-ray emission, and showed a broad continuum range of optical time lags interpreted as continuous dissipation of a reservoir of energy[5]. The declining emission implies that the jet was probably not at peak at this time. On the other hand, GX 339-4 displayed a clear ~0.1 s optical delay[8], but no corresponding state transition with changing jet activity was probed in that case. Our observations of such changes in V404 Cygni firmly support the association with inner jet activity argued for GX 339-4. This is true despite differences between the two systems. In particular, V404 Cygni has an orbital period almost 4 times longer than GX 339-4 and thus a much larger accretion reservoir. This large reservoir may be responsible for the pronounced slower (~100-1000 s long) variations exclusive to V404 Cygni, and interpreted either as instabilities in the disc[28] or as an extended corona (Dallilar et al. submitted). Irrespective, the sub-second flaring can be cleanly separated from these slower fluctuations in terms of colour and characteristic timescales[15], implying an independent origin.

1. Blandford, R.D. & Znajek, R.L. Electromagnetic extraction of energy from Kerr black holes. *Mon. Not. R. Astron. Soc.* **179**, 433—456 (1977).
2. Blandford, R.D. & Payne, D.G. Hydrodynamic flows from accretion discs and the production of radio jets. *Mon. Not. R. Astron. Soc.* **199**, 883—903 (1982).
3. Meier, D.L., Koide, S. & Uchida, Y. Magnetohydrodynamic production of relativistic jets. *Science* **291**, 84—92 (2001).
4. Merloni, A. et al. Magnetic flares and the optical variability of the X-ray transient XTE J1118+480. *Mon. Not. R. Astron. Soc.* **318**, L15—L19 (2000).
5. Malzac, J. et al. Jet-disc coupling through a common energy reservoir in the black hole XTE J1118+480. *Mon. Not. R. Astron. Soc.* **351**, 253—264 (2004).
6. Yuan, F. et al. An Accretion-Jet Model for Black Hole Binaries: Interpreting the Spectral and Timing Features of XTE J1118+480. *Astrophys. J.* **620**, 905—914 (2005).
7. Veledina, A. et al. Hot accretion flow in black hole binaries: a link connecting X-rays to the infrared. *Mon. Not. R. Astron. Soc.* **430**, 3196—3212 (2013).
8. Gandhi, P. et al. Rapid optical and X-ray timing observations of GX 339-4: flux correlations at the onset of a low/hard state. *Mon. Not. R. Astron. Soc.* **390**, L29—L33 (2008).
9. Markoff, S. et al. A jet model for the broadband spectrum of XTE J1118+480. Synchrotron emission from radio to X-rays in the Low/Hard spectral state. *Astron. Astrophys.* **372**, L25—L28 (2001).

10. Malzac, J. The spectral energy distribution of compact jets powered by internal shocks. *Mon. Not. R. Astron. Soc.* **443**, 299—317 (2014).
11. Polko, P. et al. Linking accretion flow and particle acceleration in jets – II. Self-similar jet models with full relativistic MHD gravitational mass. *Mon. Not. R. Astron. Soc.* **438**, 959—970 (2014).
12. Marscher, A. et al. The inner jet of an active galactic nucleus as revealed by a radio-to-gamma-ray outburst. *Nature* **452**, 966—969 (2008).
13. Cohen, M.H. et al. Studies of the jet in Bl Lacertae. I. Recollimation Shock and Moving Emission Features. *Astrophys. J.* **787**, 151—160 (2014).
14. Casella, P. et al. Fast infrared variability from a relativistic jet in GX 339-4. *Mon. Not. R. Astron. Soc.* **404**, L21—L25 (2010).
15. Gandhi P. et al. Furiously fast and red: sub-second optical flaring in V404 Cyg during the 2015 outburst peak. *Mon. Not. R. Astron. Soc.* **459**, 554—572 (2016).
16. Walton, D.J. et al. Living on a flare: Relativistic Reflection in V404 Cyg Observed by NuSTAR During its Summer 2015 Outburst. *Astrophys. J.* **839**, 110—132 (2017).
17. Loh, A. et al. High-energy gamma-ray observations of the accreting black hole V404 Cygni during its 2015 June outburst. *Mon. Not. R. Astron. Soc.* **462**, L111—L115 (2016).
18. Gandhi, P. et al. Correlated Optical and X-ray variability in V404 Cyg. *Astron. Telegr.* **7727** (2015).
19. Rodriguez, J. et al. Correlated optical, X-ray, and γ-ray flaring activity seen with INTEGRAL during the 2015 outburst of V404 Cygni. *Astron. & Astrophys*. **581**, L9—L13 (2015).
20. Durant, M. et al. High time resolution optical/X-ray cross-correlations for X-ray binaries: anticorrelation and rapid variability. *Mon. Not. R. Astron. Soc.* **410**, 2329—2338 (2011).
21. Khargharia, J., Froning, C. S. & Robinson, E. L., Near-infrared Spectroscopy of Low-mass X-ray Binaries: Accretion Disk Contamination and Compact Object Mass Determination in V404 Cyg and Cen X-4. *Astrophys. J.* **716**, 1105—1117 (2010).
22. Shahbaz, T. et al. Evidence for magnetic field compression in shocks within the jet of V404 Cyg. *Mon. Not. R. Astron. Soc.* **463**, 1822—1830 (2016).
23. Motta, S. et al. The black hole binary V404 Cygni: an obscured AGN analogue. *Mon. Not. R. Astron. Soc.* **468**, 981—993 (2017).
24. Jamil, O., Fender, R.P. & Kaiser, C.R. iShocks: X-ray binary jets with an internal shock model. *Mon. Not. R. Astron. Soc.* **401**, 394—404 (2010).
25. Falcke, H., Koerding, E. & Markoff, S. A scheme to unify low power accreting black holes. *Astron. Astrophys.* **414**, 895—903 (2004).
26. Merloni, A. et al. A fundamental plane of black hole activity. *Mon. Not. R. Astron. Soc.* **345**, 1057—1076 (2003).
27. Kanbach, G. et al. Correlated fast X-ray and optical variability in the black-hole candidate XTE J1118+480. *Nature* **414**, 180—182 (2001).
28. Kimura, M. et al. Repetitive patterns in rapid optical variations in the nearby black-hole binary V404 Cygni. *Nature* **529**, 54—58 (2016).

**Corresponding author** Correspondence to P.G.

**Acknowledgements** This research has made use of data from the *NuSTAR* mission, a project led by the California Institute of Technology, managed by the Jet Propulsion Laboratory, and funded by the National Aeronautics and Space Administration. We thank the *NuSTAR* Operations, Software, and Calibration teams for support with the execution and analysis of these observations. This research has made use of the *NuSTAR* Data Analysis Software (NuSTARDAS) jointly developed by the ASI Science Data Center (ASDC, Italy) and the California Institute of Technology (USA), as well as the high energy astrophysics science archive (HEASARC). P.G. thanks STFC for support (grant reference ST/J003697/2). ULTRACAM and V.S.D. are supported by STFC grant ST/M001350/1. P.G. thanks C.B. Markwardt, C.M. Boon, M. Fiocchi, K. Forster, A. Zoghbi, and T. Muñoz-Darias for help and useful discussions. J.C. acknowledges financial support from the Spanish Ministry of Economy, Industry and Competitiveness (MINECO) under the 2015 Severo Ochoa Program MINECO SEV-2015-0548, and to the Leverhulme Trust through grant VP2-2015-04. T.R.M. acknowledges STFC (ST/L000733/1). J.M. acknowledges financial support from the french National Research Agency (CHAOS project ANR-12-BS05-0009), and D.A. thanks the Royal Society. S.M. acknowledges support from NWO VICI grant Nr. 639.043.513. Patrick Wallace is gratefully acknowledged for use of his SLA C library. P.A.C. is grateful to the Leverhulme Trust for the award of a Leverhulme Emeritus Fellowship. Part of this research was supported by the UK-India UKIERI/UGC Thematic Partnership grants UGC 2014-15/02 and IND/CONT/E/14-15/355. We thank the reviewers for suggestions that helped to make the discussion and interpretation more robust. This work profited from discussions carried out during a meeting organised at the International Space Science Institute (ISSI) Beijing by T. Belloni and D. Bhattacharya.

**Author Contributions** P.G. wrote the ULTRACAM proposal, analysed the data and wrote the paper. The ULTRACAM observations were coordinated and carried out by L.K.H., S.P.L., V.S.D. and T.R.M. The X-ray observations were proposed by D.J.W., coordinated by D.S., J.A.T. and F.A.H., and the timing data analysed by M.B. Radio data were obtained and analysed by R.F. and K.M. *INTEGRAL* data was arranged by E.K. The remaining authors provided insight into jet physics constraints (S.M., J.M., C.C.), cross-correlation analyses (P.C., R.I.H., C.K., C.F., M.P., F.V.), and placing the source in context (D.A., J.C., P.A.C., D.M.R., F.R., A.S.). All authors read and commented on multiple versions of the manuscript.

**Author Information**
*Corresponding author. [1]Department of Physics and Astronomy, University of Southampton, SO17 3RT, UK. [2]INAF-Osservatorio Astronomico di Cagliari, via della Scienza 5, I-09047 Selargius, Italy. [3]Department of Physics and Astronomy, University of Sheffield, Sheffield S3 7RH. [4]Instituto de Astrofisica de Canarias, 38205 La Laguna, Santa Cruz de Tenerife, Spain. [5]Astrophysics, Department of Physics, University of Oxford, Keble Road, Oxford OX1 3RH, UK. [6]Cahill Center for Astrophysics, 1216 East California Boulevard, California Institute of Technology, Pasadena, CA 91125, USA. [7]IRAP, Université de toulouse, CNRS, UPS, CNES, Toulouse, France. [8]Anton Pannekoek Institute for Astronomy, University of Amsterdam, 1098 XH Amsterdam, The Netherlands. [9]Deptt of Physics, University of Warwick, Gibbet Hill Road, Coventry, CV4 7AL, UK. [10]Jet Propulsion Laboratory, California Institute of Technology, 4800 Oak Grove Drive, Mail Stop 169-221, Pasadena, CA 91109, USA. [11]Space Sciences Laboratory, 7 Gauss Way, University of California, Berkeley, CA 94720-7450, USA. [12]Institute of Astronomy, University of Cambridge, Madingley Road, Cambridge CB3 0HA. [13]INAF-Osservatorio Astronomico di Roma, Via Frascati 33, I-00040 Monteporzio Catone, Italy. [14]DiSAT, Universitá degli Studi dell'Insubria, Via Valleggio 11,I–22100 Como, Italy. [15]INAF - Osservatorio Astronomico di Brera Merate, via E. Bianchi 46, I-23807 Merate, Italy. [16]Departmento de Astrofísica Universidad de La Laguna (ULL), E-38206 La Laguna, Tenerife, Spain. [17]ISDC, Department of Astronomy, University of Geneva, Chemin d'Ecogia 16, CH-1290 Versoix, Switzerland. [18]Department of Physics and Astronomy, Louisiana State University, 202 Nicholson Hall, Tower Drive, Baton Rouge, LA 70803, USA. [19]ESA/ESTEC, Keplerlaan 1, 2201 AZ Noordwijk, The Netherlands. [20]Inter-University Centre for Astronomy and Astrophysics, Post Bag 4, Ganeshkhind, Pune-411007, India. [21]European Southern Observatory, K. Schwarzschild-Strasse 2, D-85748 Garching bei München, Germany. [22]New York University

Abu Dhabi, PO Box 129188, Abu Dhabi, UAE. [23]Department of Physics, University of Alberta, CCIS 4-183, Edmonton, AB T6G 2E1, Canada.

**Competing interests** The authors declare no competing financial interests.

# Methods

**Observations. (1) ULTRACAM.** ULTRACAM[29] is a fast optical imager capable of high sampling rates (up to several hundred Hz) in three filters simultaneously. The ULTRACAM data on the 2015 outburst of V404 Cygni, without any X-ray analyses, have been described in ref. 15, focusing on June 26, the day following the light curve described herein. The time resolution ($\Delta T$) during the June 25 observation used here was 35.94 ms, constant throughout the observation. This is the cycle time between consecutive exposures, and includes a short dead time of approximately 1 ms. Data were obtained in three optical filters simultaneously ($u'$, $g'$ and $r'$), but the $r'$ data had the best combination of signal-to-noise ratio (SNR) and time resolution, by far. The SNR is a factor of 2 better in $r'$ than $g'$ and the time resolution is a factor of 15 worse in $u'$ because of the need for coadding frames to gain sensitivity. Hence, we focus on the $r'$ data in this work, though the source behaviour is qualitatively similar in all optical filters. The light curve times were transformed to the Solar system barycentre by using custom ULTRACAM software, based upon the positional astronomy library SLALIB.

**(2) *NuSTAR*.** *NuSTAR* (the *Nuclear Spectroscopic Telescope Array*)[30] is the first X-ray telescope in orbit capable of focusing photons above 10 keV and is highly sensitive up to approximately 80 keV. *NuSTAR* carried out a long (∼3 day) Target of Opportunity observation triggered by the outburst of V 404 Cygni, and we were able to coordinate a ∼2 hour optical observation with WHT/ULTRACAM in the early morning hours of 2015 June 25, which is the focus of the analysis herein.

The *NuSTAR* data are described in more detail by Walton et al.[16]. Here we give salient details only. The *NuSTAR* detectors are not affected by pileup issues due to finite read times. But at high count rates above several hundred per second, dead time becomes important, and the X-ray light curves herein were all corrected for dead time[31]. The full energy range of 3—79 keV is used, summing up counts from both focal plane modules (FPMA and FPMB). Events were barycentered using BARYCORR with FTOOLS[32] with a clock correction file that covered the full observation. For cross-correlating the multiwavelength light curves, the X-ray data were sampled using identical time bins to the optical. In order to do this, the X-ray barycentered event times were converted to Universal Time, Coordinated (UTC) and then binned using the exact time bin boundaries (which are uniformly spaced in UTC). The binned X-ray light curves then exactly match the barycentred optical data. Background is ignored as the source completely dominates the *NuSTAR* field of view.

**(3) Epochs of simultaneity.** After removing the Earth occultation gap and a few data points at the edges of the gap, there are two epochs of strict simultaneity between optical and X-rays. Epoch 1 is 2112.6 s long, starting at Barycentric Modified Julian Date (BMJD) 57198.15404876 and finishing at 57198.17850000. Epoch 2 has a duration of 1820.3 s, and lasts between BMJD 57198.20800000 & 57198.22906829.

**(4) AMI.** Radio imaging data were obtained by the Arcminute Microkelvin Imager (AMI)[33]. Those data will be described in Fender et al. (in prep.). For the analysis here, we use light curves extracted using time bins of 20 s at 13.9 and 15.4 GHz. The light curve in Fig. 1 shows a brightening in radio flux by a factor of approximately 5 in Epoch 2, as compared to Epoch 1. In addition, the radio spectral index also rises dramatically, from being significantly negative in Epoch 1, to being consistently above 0 in Epoch 2, with a maximum value of 0.68±0.17. A factor of 15 binning was employed to increase the signal-to-noise of the spectral index trend.

**(5) *INTEGRAL*.** The *INTEGRAL*[34] mission provided some of the best coverage of the 2015 outburst. Light curve data products in standard bands for all instruments have been made publicly available[35]. The IBIS/ISGRI instrument[36] light curve covering the energy range of 25—200 keV is displayed in Fig. 2 to show the long term evolution starting from before our ULTRACAM/*NuSTAR* campaign and continuing until outburst peak on June 26. The light curve demonstrates that the state transition we caught was unique during this period, with the source remaining bright and flaring until the end, even after the *NuSTAR* observation had finished.

**Timing correlation and lag**

One popular estimator of the degree of correlation between two bands is the discrete correlation function (DCF), which measures the averaged cross product between all pairs of lags defined by the sampling times of the two light curves. We computed the DCF between the optical and the X-rays according to the algorithm of Edelson & Krolik (1991)[37], for the strictly simultaneous epochs separately. Each epoch was divided into $10^3$ independent segments, and the DCF computed over time lags of ±1 s after pre-whitening to remove any red noise trend[38]. The mean DCF and its standard error were computed at each time lag, and the results for the two epochs are shown in Fig. 1.

We quantified the lags by computing their centroid, which is defined as the mean of the individual lags ($t_i$) weighted by the DCF coefficients $c_i$ (refs. 39, 40):

$$\tau = \sum_i (c_i \times t_i) / \sum_i (c_i)$$

Following standard practice[40], the summation is over all $t_i$ where the DCF coefficients $c_i$ have values above half of the peak (representing an approximate weighted mean of the full-width DCF at half-maximum). The error on $\tau$ is computed by generating an ensemble of $10^4$ DCFs by using the scatter in the coefficients from the independent data segments to randomise the coefficients at all lags. This results in $10^4$ mean-weighted centroid lags ($\tau_{\text{random}}$), and the standard deviation of $\tau_{\text{random}}$ is used as the error estimator of $\tau$.

**Data Availability Statement**

The data that support the plots within this paper and other findings of this study are available from public archives and the corresponding author upon reasonable request.

29. Dhillon, V. S. et al. ULTRACAM: an ultrafast, triple-beam CCD camera for high-speed astrophysics. *Mon. Not. R. Astron. Soc.* **378**, 825—840 (2007).
30. Harrison, F. et al. The *Nuclear Spectroscopic Telescope Array* (*NuSTAR*) High-energy X-Ray Mission. *Astrophy. J.* **770**, 103—131 (2013).
31. Bachetti, M. et al. No Time for Dead Time: Timing Analysis of Bright Black Hole Binaries with NuSTAR. *Astrophys. J.* **800**, 109—120 (2015).

32. Blackburn, J. K. FTOOLS: A FITS Data Processing and Analysis Software Package. *Astr. Soc. P.* **77**, 367—370 (1995).
33. Zwart, J. T. L. et al. The Arcminute Microkelvin Imager. *Mon. Not. R. Astron. Soc.* **391**, 1545—1558 (2008).
34. Winkler, C. et al. The *INTEGRAL* mission. *Astron. & Astrophys.* **411**, L1—L6 (2003).
35. Kuulkers, E. INTEGRAL observations of V404 Cyg (GS 2023+338): public data products. *Astron. Telegr.* **7758** (2015).
36. Ubertini, P. et al. IBIS: The Imager on-board INTEGRAL. *Astron. Astrophys.* **411**, L131—L139 (2003).
37. Edelson, R. A. & Krolik, J. H. The discrete correlation function - A new method for analyzing unevenly sampled variability data. *Astrophys. J.* **333**, 646—659 (1988).
38. Welsh, W.F. On the Reliability of Cross-Correlation Function Lag Determinations in Active Galactic Nuclei. *Publ. Astron. Soc. Pac.* **111**, 1347—1366 (1999).
39. Koratkar, A. P. & Gaskell, C. M. Structure and kinematics of the broad-line regions in active galaxies from IUE variability data. *Astrophys. J. Supp.* **75**, 719—750 (1991).
40. Peterson, B. M. et al. On Uncertainties in Cross-Correlation Lags and the Reality of Wavelength-dependent Continuum Lags in Active Galactic Nuclei. *Pub. Astron. Soc. Pacific.* **110**, 660—670 (1998).

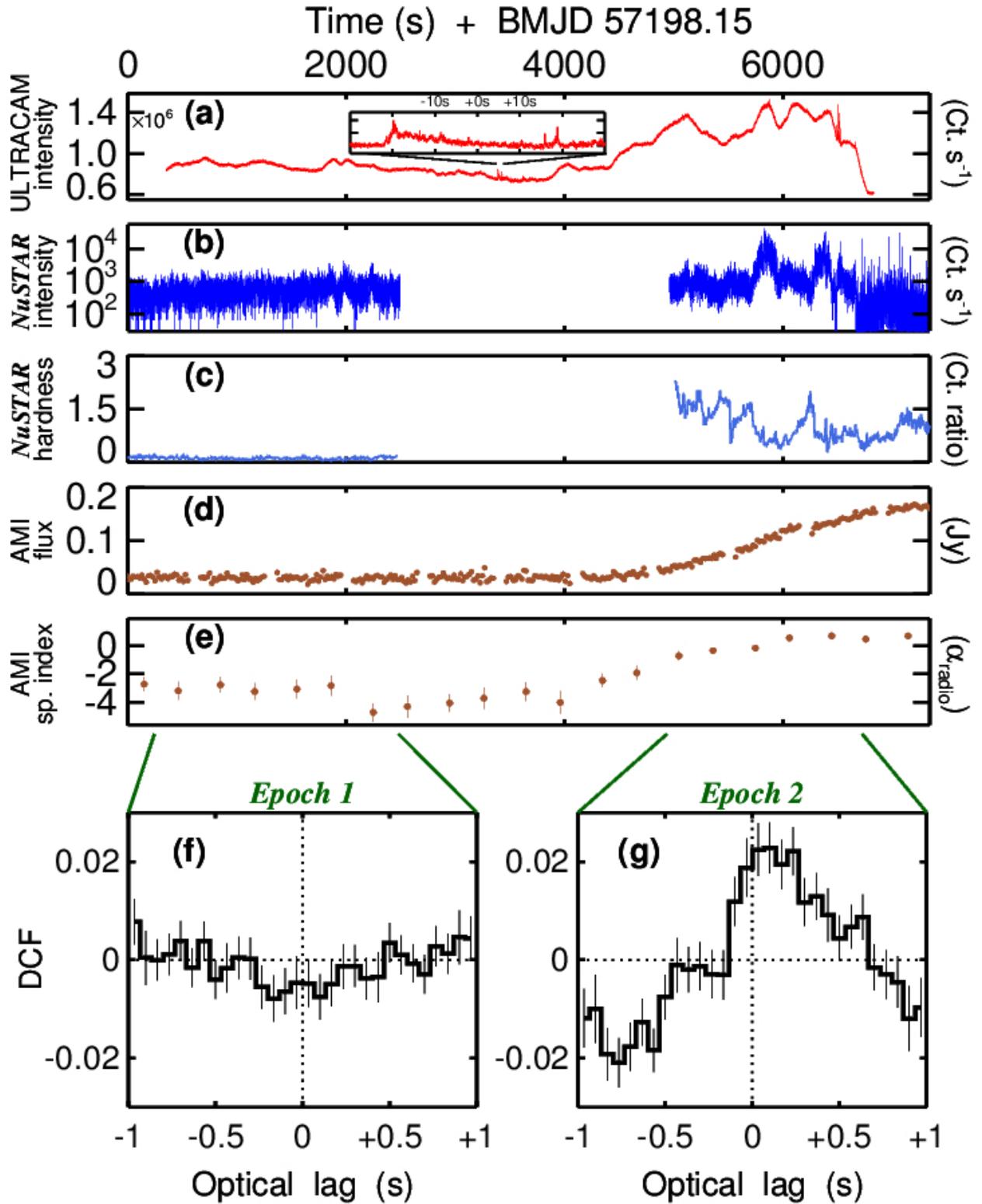

**Figure 1: Multiwavelength light curves and timing correlations of V404 Cygni on 2015 June 25.** All times have been transformed to the common Solar System Barycentre (BMJD). (a) WHT/ULTRACAM optical *r'* band light curve at a time resolution of 35.94 ms. The inset zoom-in highlights a short fast flaring episode, likely marking the moment of the serendipitously-caught state transition. (b) *NuSTAR* full band X-ray light curve. There are two epochs of strict simultaneity with the optical, corresponding to two satellite orbits, separated by an Earth occultation gap. (c) *NuSTAR* hardness ratio (HR) between bands 10-79 keV and 3-10 keV, lightly binned with a 5 s kernel for display purposes. (d) AMI 15.4 GHz light curve. Uncertainties for panels (a) to (d) are typically smaller than symbol sizes. (e) Radio spectral index between 15.4 GHz and 13.9 GHz ($F_\nu \propto \nu^{\alpha_{radio}}$) with errors denoting 1 std. dev. over the ≈350 s-long segments used for binning. Panels (f) and (g): Optical vs. X-ray discrete cross-correlation functions (DCF) between lag times of -1 to +1 s, for the two epochs, respectively. Uncertainties denote 1 std. error on the mean for each lag bin. A positive delay corresponds to the optical lagging the X-rays.

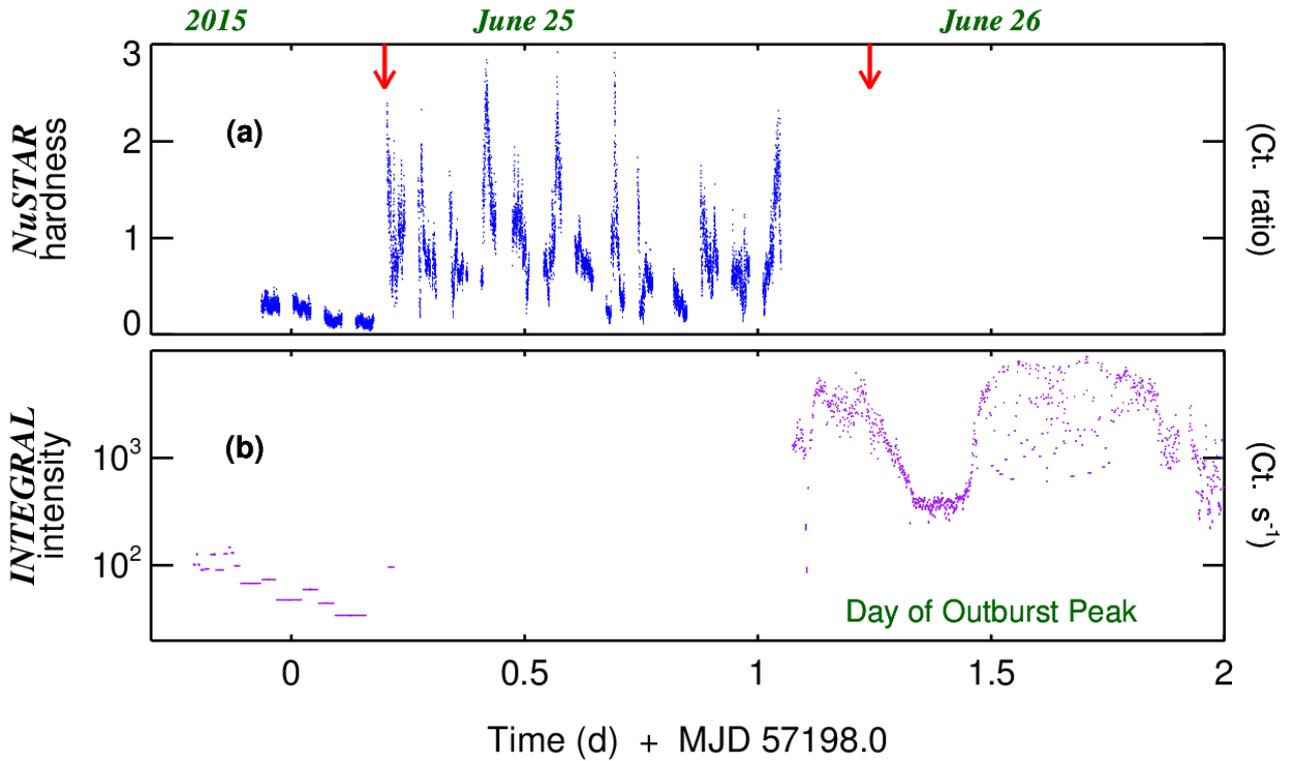

**Figure 2: <u>X-ray evolution of V404 Cygni leading to outburst peak.</u>** (a) The full *NuSTAR* hardness ratio light curve between 10-79 keV and 3-10 keV, plotted using 5 s time bins. Uncertainties (1 std. dev.) are typically less than 10% on any point and are omitted for clarity of display. (b) *INTEGRAL* IBIS/ISGRI 25-200 keV light curve covering the period of 2015 June 24-26. Uncertainties denoting 1 std. dev. are plotted but are typically smaller than symbol sizes. The peak of the outburst occurred on June 26, following which the outburst gradually declined to quiescence. The red arrows in panel (a) denote the times of ULTRACAM optical observations on June 25 covering the state transition (see Fig. 1) and June 26 (when rapid optical flaring was highly developed; ref. 15). Bin widths denote the time interval of data averaging.

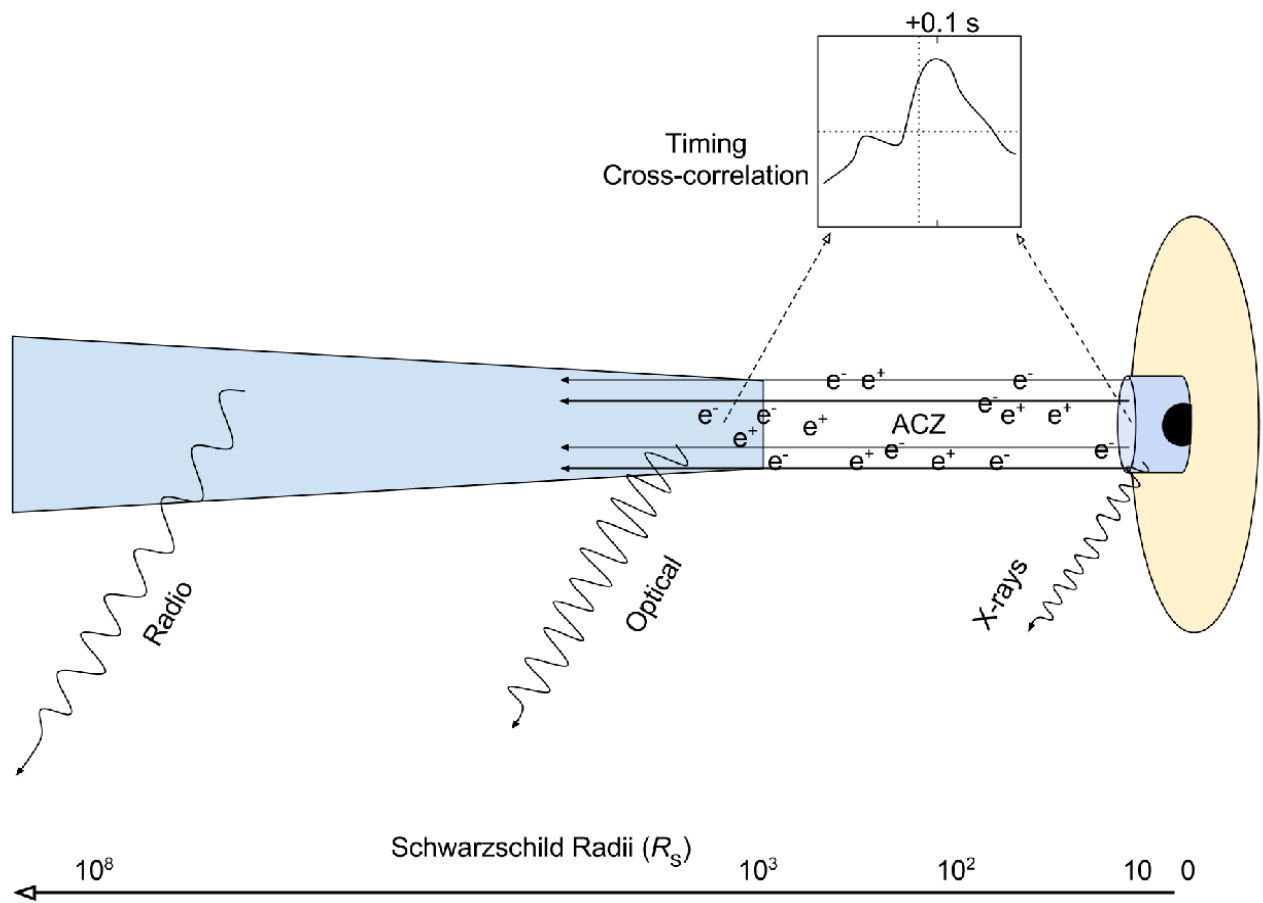

**Figure 3: <u>Schematic of the post-transition accretion and jet geometry of V404 Cygni</u>.** The black hole and accretion disc are situated off to the right. X-rays originate in a compact region within ~5 $R_S$ of the black hole due to Comptonisation (either jet or coronal). Optical photons are delayed with respect to X-rays by ~0.1 s. This 'optical base' lies ≤$10^3$ $R_S$ from the X-ray core, and should power broadband synchrotron radiation extending to high frequencies. The above time delay limits the extension of the putative acceleration and collimation zone (ACZ), which is likely to be Poynting-flux dominated. Beyond the optical base, shocks in the longitudinally and laterally expanding jet inject energy that power the broadband lower frequency radiation down to the radio.

# An elevation of 0.1 light-seconds for the optical jet base in an accreting Galactic black hole system:

# SUPPLEMENTARY INFORMATION

*P. Gandhi, M. Bachetti, V.S. Dhillon, R.P. Fender, L.K. Hardy, F.A. Harrison, S.P. Littlefair,
J. Malzac, S. Markoff, T.R. Marsh, K. Mooley, D. Stern, J.A. Tomsick, D.J. Walton,
P. Casella, F. Vincentelli, D. Altamirano, J. Casares, C. Ceccobello, P.A. Charles,
C. Ferrigno, R.I. Hynes, C. Knigge, E. Kuulkers, M. Pahari, F. Rahoui,
D.M. Russell, A.W. Shaw*

**Spectral energy distributions**

The broadband average spectral energy distributions (SEDs) for Epochs 1 and 2 are shown in Supplementary Fig. 1.

In the optical, we plot dereddened fluxes for both epochs, as well as the full range in fluxes resulting from source variability. The reddening toward the source remains controversial. In a recent measurement based upon highly polarised field stars surrounding V404 Cygni, the extinction has been constrained to be $3.0 < A_V$ (mag) $< 3.6$, or a mean value of $A_V=3.3$ mag[1]. This is lower than the previous measurement by Casares et al.[2] of $A_V=4.0$ mag, although it is formally consistent within an assigned uncertainty of 10%. Part of the uncertainty results from ambiguity in the exact donor star spectral type (discussed below). Another recent estimate is $A_V=3.82\pm0.36$ mag, based upon the equivalent width of a diffuse interstellar band feature[3]. Given this spread of values, we use an intermediate $Av=3.5$ mag to deredden the observed ULTRACAM fluxes, together with a standard Galactic reddening law[4].

The optical fluxes rises towards redder filters, as is characteristic of optically-thin synchrotron emission. However, the propagated uncertainty from the above quoted measurements is $\Delta Av=0.6$ mag. All uncertainties in this manuscript are quoted at 68% confidence (1 std. dev.) unless otherwise stated. Such a large systematic extinction uncertainty precludes definitive statements about the absolute values of the spectral slopes in the optical/ultraviolet range. In addition, we suggest that caution must be employed when claiming broadband spectral breaks in the optical/ultraviolet without other corroborating evidence. On the other hand, the fact that the sub-second flaring is significantly *redder* than the slow fluctuations that dominate the overall SED is unambiguous[5].

Our focus here is timing analysis and we defer to ref. 6 for the spectral analysis of the *NuSTAR* data. For the broadband SEDs presented in Supplementary Fig. 1, we carried out simple parametric fits to the observed data with an absorbed powerlaw and reflection component, including a blurred Fe line. For clarity, only the FPMA data are plotted. The difference between the two epochs is striking, with a much harder spectrum in Epoch 2 and a jump in observed flux by a factor of ~15. No absorption correction is performed since the source spectra are highly complex[6, 7].

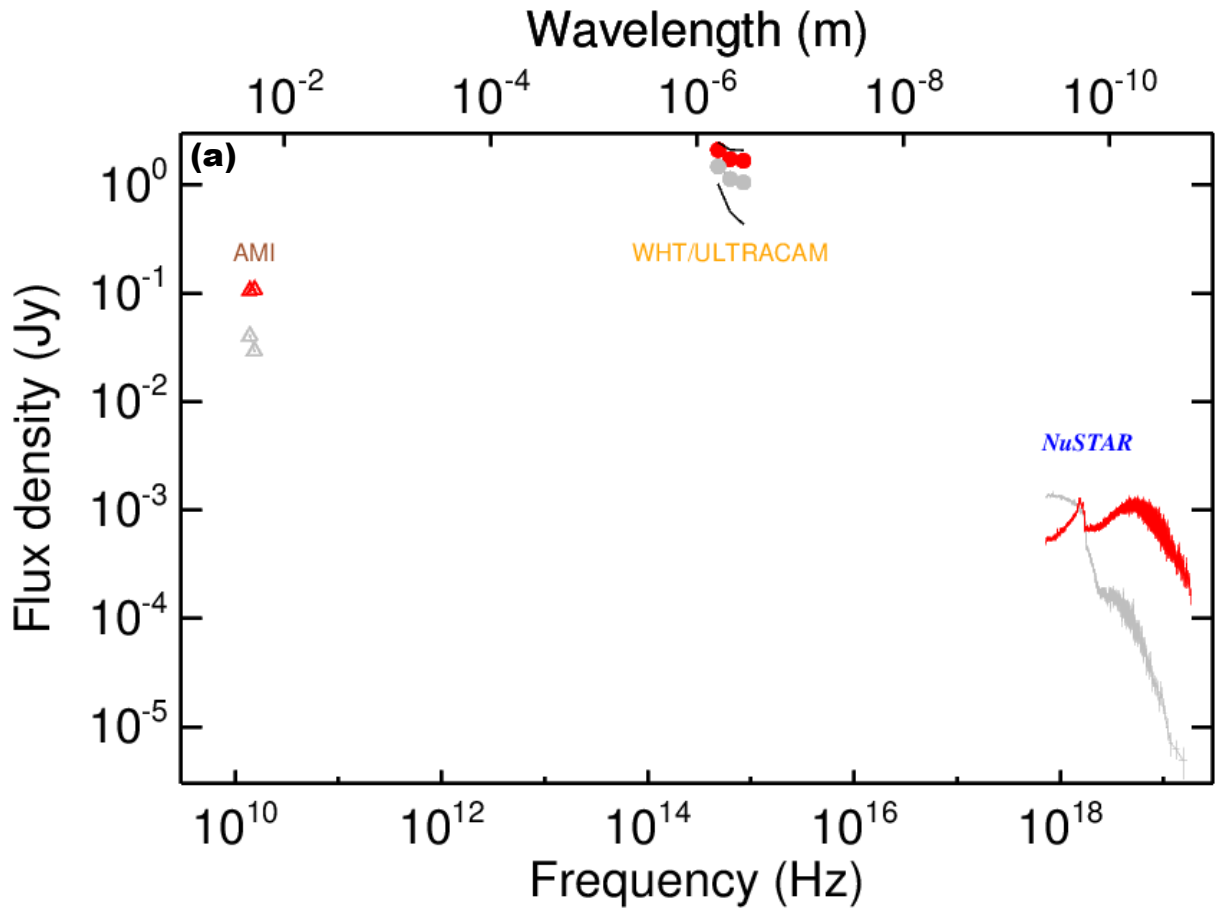

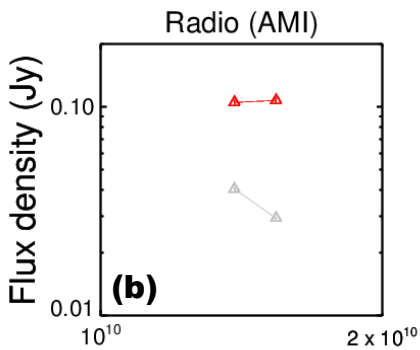
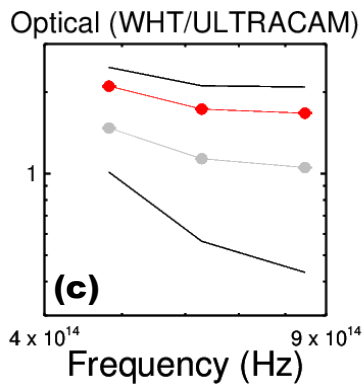
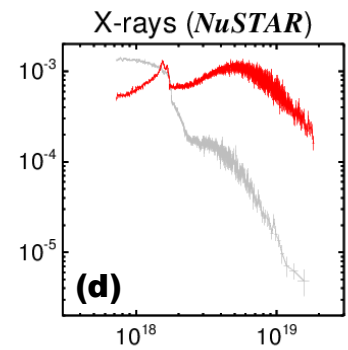

**Supplementary Figure 1: <u>Broadband spectra during Epochs 1 and 2.</u> (a)** Strictly simultaneous, dereddened, mean spectral energy distributions (SEDs) of V404 Cygni on 2015 June 25. (Bottom panels) Zoom-ins on the three bands of Radio (b), Optical (c) and X-rays (d). In all panels, the lower grey and upper red sets of SEDs correspond to Epochs 1 and 2 (pre- and post-transition), respectively. Two black curves enveloping the optical data represent the maximal variability over the duration of the observation. The X-ray spectra are modelled using phenomenological power-law, reflection and fluorescence components, simply meant to accurately represent the observed flux. Uncertainties denote 1 mean std. error on each bin.

**Fourier domain coherence and lags**

Supplementary Fig. 2 shows the frequency domain correlations, in terms of the optical/X-ray coherence, phase lags and time lags. The coherence and the lags represent the relative magnitude and the phase angle of the complex-valued cross spectrum, respectively. These are computed according to a canonical recipe[8] over independent light curve segments, which are then averaged. As a balance between the need for several segments and the relatively short light curve duration, here we used 8 segments containing 8192 bins (~294 s in duration each). As recommended[8], the coherence and lags can be usefully computed in the cases where the power spectral densities and coherence values are significantly above the expected noise threshold at any given Fourier frequency, and we use a threshold of 3 times greater than the noise here. For a more detailed description, we refer the reader a similar optical/X-ray analysis on GX 339-4 presented in ref. 9, and also note that a generalisation of the algorithm will soon be presented by Vincentelli et al. (in prep.).

Tests on short light curve segments showed the presence of non-stationary behaviour in the cross-correlations, but there are two sets of signals with significant coherence that stand out across the full data set. Firstly, the coherence peaks at low Fourier frequencies at and below ~0.01 Hz, corresponding to the 'slow variations' reported in ref. 5. The optical in this case shows long (but variable) lags with respect to X-rays, ranging over ~tens of seconds[10]. The broadband optical colours of these slow variations are distinct from the fast flaring under consideration here, pointing to a distinct origin.

The other correlated signal appears at high Fourier frequencies during Epoch 2. At ~0.5—1 Hz, the coherence is significantly above the noise threshold, and the corresponding optical lags for these fluctuations are close to ~+0.1 s. At ≈1.5 Hz, the time lag is most significant, $\tau_{1.5Hz}$=0.131±0.028 s, consistent with the weighted DCF lag that we focus on in the main text (Fig. 1). Any corresponding lag in Epoch 1 is associated with significantly weaker inter-band coherence, consistent with the absence of any clear signal in the time-domain DCF (Fig. 1). Finally, we note that the optical never clearly leads the X-rays, i.e. there are no significant negative phase lags, arguing against physical processes such as synchrotron self-Compton with optical photons seeding (and thus leading) X-ray emission.

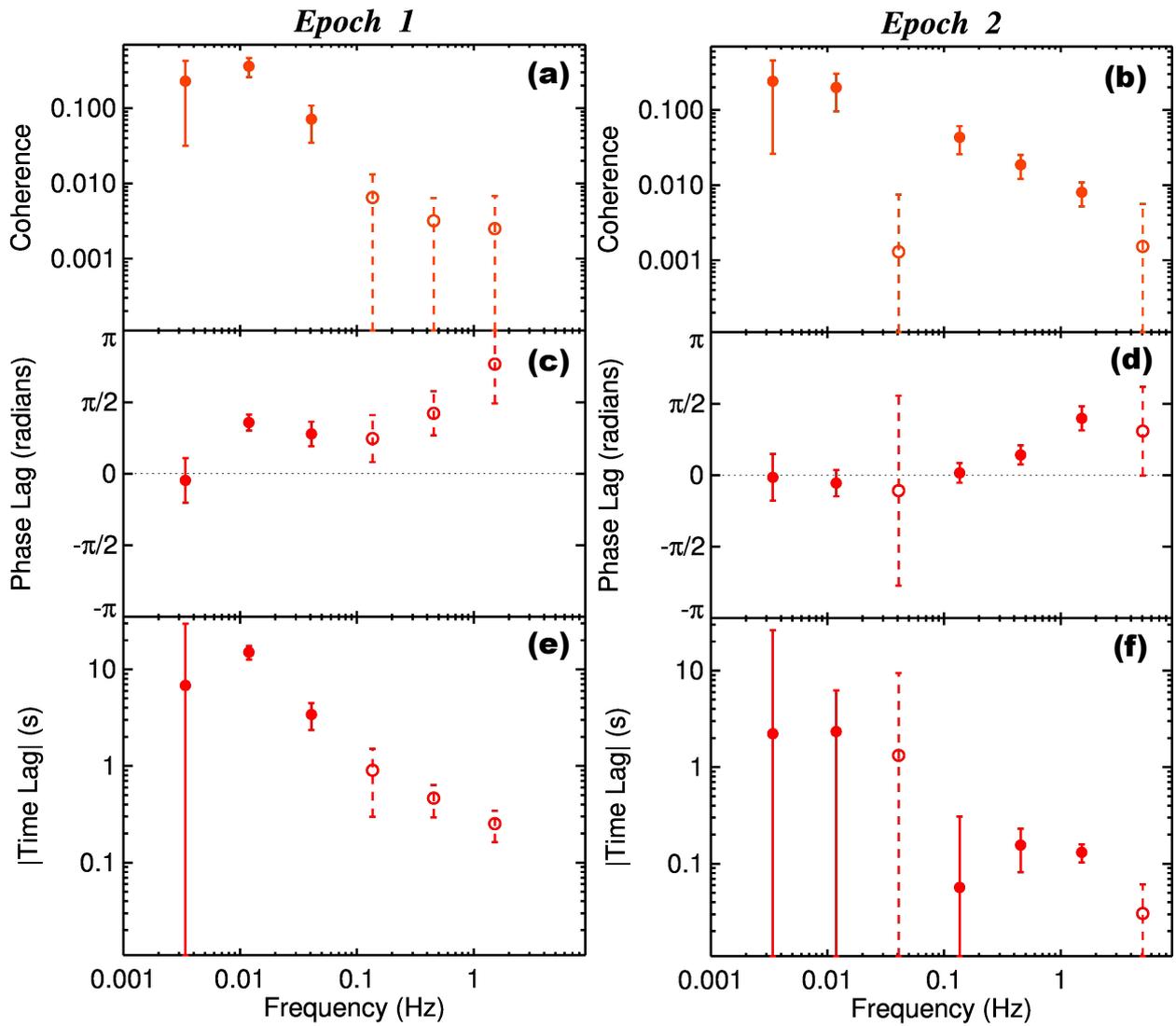

**Supplementary Figure 2: <u>Fourier Coherence and Lags.</u>** Coherence (panels a, b), phase lags (c, d) and optical time lags (e, f) for the simultaneous optical and X-ray light curve epochs. Frequency bins where the coherence is significantly above noise are plotted using filled symbols with thick continuous lines for error bars. Bins where the coherence was below the threshold are plotted with empty symbols and dashed error bars. There are no bins where the phase lags are significantly negative. In the logarithmic time lag panels, the lags are plotted as absolute values. Uncertainties denote 1 mean std. error at each frequency bin.

**Null hypothesis simulations for presence of a correlation**

We tested the probability of a false correlation by using simulations of uncorrelated light curves. For this purpose, we generated an ensemble of $10^4$ stochastic optical light curves, each time randomising the phases of the observed ULTRACAM Epoch 2 light curve. Each randomisation produced a light curve that preserved the variability characteristics of the data (i.e. the power spectrum of the simulated light curve is identical to that of the observed ULTRACAM data), but with arbitrary variations as a function of time. When correlated with the observed *NuSTAR* light curve, no correlation is expected for this ensemble, and the fraction of simulations that result in a DCF exceeding that seen in the data yields the null hypothesis probability. The result is shown in Supplementary Fig. 3.

We first ask what the null hypothesis probability is for the presence of any sub-second cross-correlation between the bands, without restricting ourselves to any particular lag value. Over the full computed DCF range of -1 to +1 s, 93 simulated DCFs exceed the absolute peak value of the observed DCF, i.e. a null hypothesis probability (percentage) of 0.93 % for a lag arising by chance. Randomising the high frequency phases alone (i.e. preserving the low frequency long lags in the simulated light curves makes little difference to the results.

A more restricted question is to ask about the significance of the identified weighted lag feature at $\tau=+0.13$ s. Only 2 of the $10^4$ simulated DCFs lie above the Epoch 2 DCF in the corresponding lag bin. However, the feature has a finite width, and taking into account the uncertainty $\Delta\tau=0.04$ s, adjacent bins need to be included, resulting in a corresponding null hypothesis probability of 0.16 %.

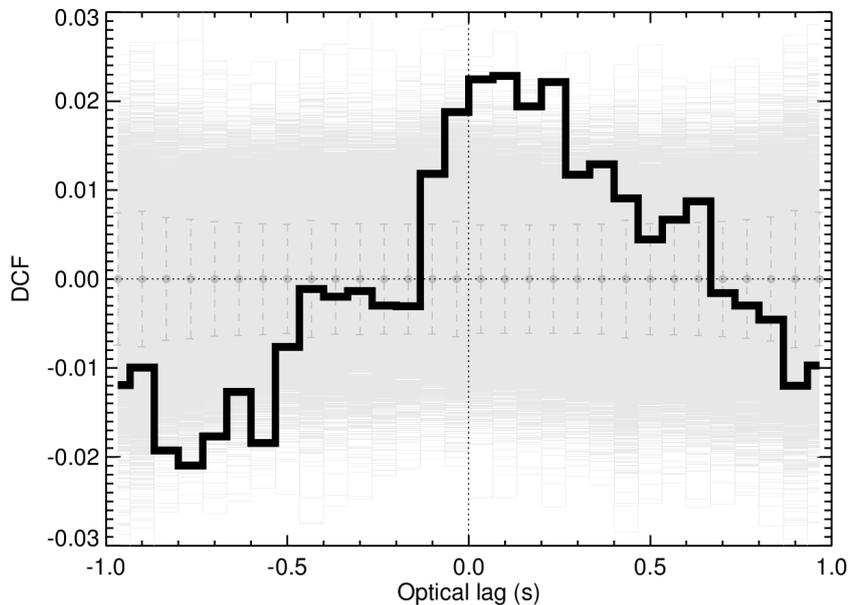

**Supplementary Figure 3: <u>Simulations of uncorrelated light curves</u>. In grey are shown 10,000 simulated DCFs, and overplotted in thick black is the data DCF (for uncertainties on this, see Fig. 1f). For each simulation, the phases of the observed Epoch 2 optical light curve were randomised, and this randomised light curve was then cross-correlated with the observed Epoch 2 X-ray light curve. No correlation is expected in this case, and the fraction of simulations that lie above the data yields the null hypothesis probability of the observed Epoch 2 correlation arising by chance. The thick dashed grey points with error bars denote the mean value of the 10,000 simulated DCFs at any given lag and 1 std. dev., respectively.**

## The correlation in shorter sub-segments of data

As a final check, we simply split both epochs 1 and 2 into two equal halves, and show the result in Supplementary Fig. 4. While the lag is clearly preserved in both independent halves of Epoch 2, there is no (or only a very weak lag) in Epoch 1. For Epoch 2, the weighted mean optical lag with respect to X-rays is $\tau=+0.09\pm0.04$ s and $\tau=+0.15\pm0.05$ s, for the first and second halves, respectively. These are consistent (at 2 std. dev.) with the mean weighted lag for the full Epoch 2 quoted in the main text of the paper. The only putative significant lag in Epoch 1 is a weak anti-correlation during the first half. We note the mean weighted (anti-correlation) lag in this case to be $\tau=-0.05\pm0.08$ s, which is not significant.

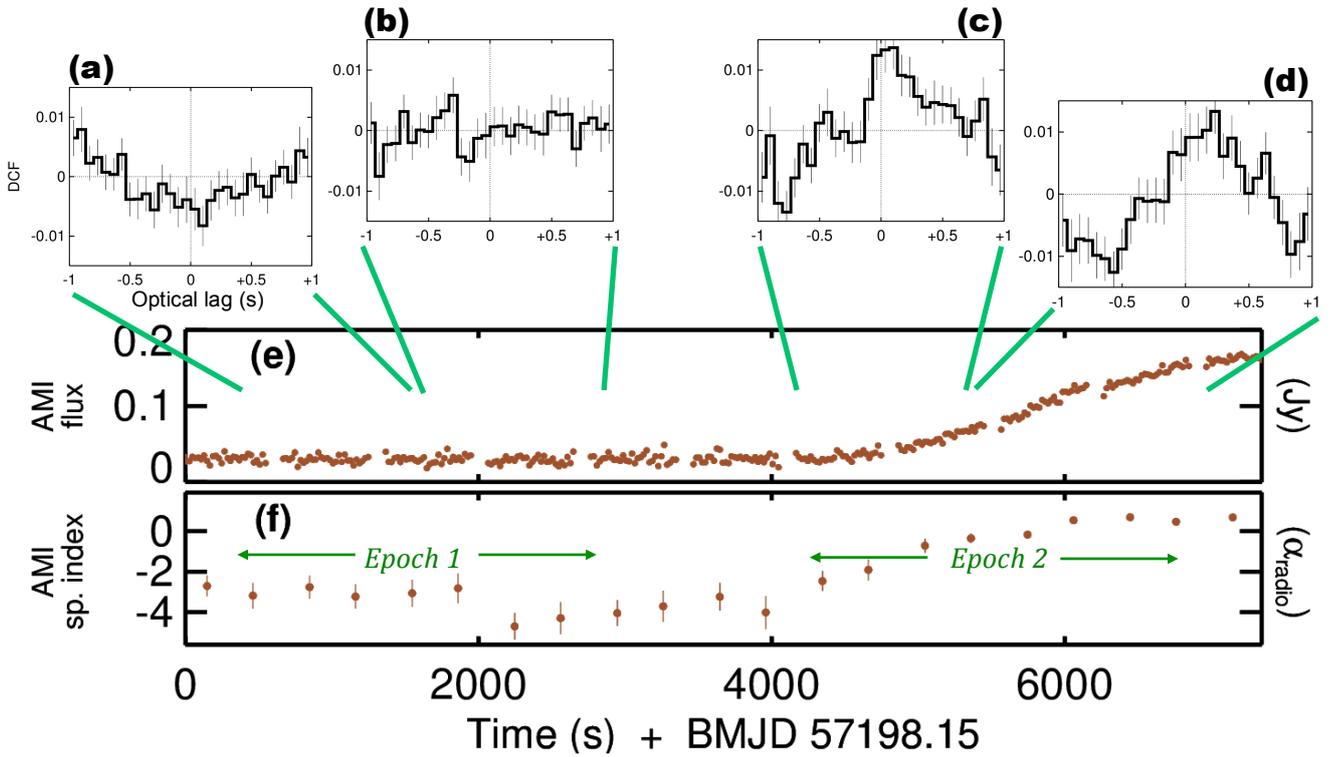

**Supplementary Figure 4: <u>Testing the DCF on sub-segments of the light curves.</u> Splitting the simultaneous epochs of observation each into two halves and computing the DCFs for each of the segments shows a consistent result to Fig. 1. The small panels at the top show the DCF as a function of optical lag for each of the epoch halves, with the axes being identical for each panel. Uncertainties denote 1 std. error for each bin. Neither segment of Epoch 1 (panels a and b) shows any obvious correlation, whereas both halves of Epoch 2 (c and d) show a positive correlation. Panels e and f show the radio data, identical to the bottom two long panels in Fig. 1, and we refer the reader to that figure for further details.**

**Power spectra**

The optical and X-ray power spectral densities (PSDs) for Epochs 1 and 2 are displayed in Supplementary Fig. 5. These are plotted in frequency x power ($\nu P_\nu$) units with standard rms$^2$ normalisation[11,12]. For the *NuSTAR* data, these are actually *cospectra* between the two focal plane modules or FPMs. The cospectrum is the real component of the cross spectrum between the two X-ray detectors[13], employed as a means to cancel dead time distortions to the PSDs. Both the optical and X-ray PSDs were computed using light curve segments of duration 512 s, and we restrict analysis for both energies to the Nyquist frequency of the optical of ≈13.5 Hz after mild frequency binning. For further details on the procedure, see ref. 13 for details on *NuSTAR* PSD extraction, and ref. 5 for details on the optical PSDs of V404 Cygni.

The X-ray PSDs show a higher normalisation in both epochs than the optical, reflecting the stronger X-ray variance. Integrating the r.m.s$^2$—normalised PSDs over any given Fourier frequency range should yield the fractional variance over that range, whose square-root then returns the fractional r.m.s. Numerically integrating the PSDs gives optical fractional r.m.s values of 0.02—0.06 and higher corresponding X-ray fractional r.m.s values of 0.12—0.13. In the case of the X-ray light curves, these values are likely to be lower limits to the *true* X-ray fractional r.m.s levels due to dead time. This effect has been extensively studied using realistic high count rate simulations, using the results of which (cf. Eq. 5 from ref. 13) we estimate the true X-ray fractional r.m.s levels to be 0.3—0.5. Epoch 2 has the higher normalisation, reflecting the increase in multiwavelength fluctuation strengths as compared to Epoch 1. The optical PSDs rise steeply toward low Fourier frequency since the optical r.m.s is dominated by smooth slow variations. We characterised the PSDs with a simple powerlaw model ($P_\nu \propto \nu^{-\alpha}$) and added a constant white noise level at high frequencies.

Both epochs are fitted well with a steep slope $\alpha \approx 2.1$ at the low frequency end. At high frequencies, there is excess power apparent around 1 Hz that is especially strong in Epoch 2, that can be characterised by a zero-centred Lorentzian[11]. The X-ray PSDs have a shallower decline with increasing frequency in both cases, with $\alpha \approx 1.6$ over most of the frequency range. Again, Epoch 2 displays a higher power than Epoch 1, especially at low frequencies (which show a break in the X-rays), corresponding to the broad and slow variations that we observe.

We note that the above fits are only meant as approximate characterisations of the broad PSD profiles to see how they compare between the epochs and between the energy bands. However, these power spectra are much steeper (dominated by long timescale fluctuations) as compared to those seen in other GX 339—4 in the typical hard state[8]. In that case, Malzac[14] used the assumption that the power spectrum of energy injection events in the jet matches the observed X-ray flicker noise PSD to show that optical time lags of order 0.1 s can be naturally obtained. With the present PSDs of V404 Cygni dominated instead by much longer timescale fluctuations, this assumption would also result in much longer time lags, inconsistent with the data. This either rules out the above assumption for internal shock models, or suggests that some process may be suppressing the high frequency flicker noise from being directly observed. It may also be related to the fact that our observations were carried out very soon after transition, when the source had probably not yet entered a 'canonical' hard state. In fact, on the following day of June 26, both the optical and the X-ray

PSDs were significantly flatter with a large rise in the strength of high frequency optical[5] and X-ray variations[15].

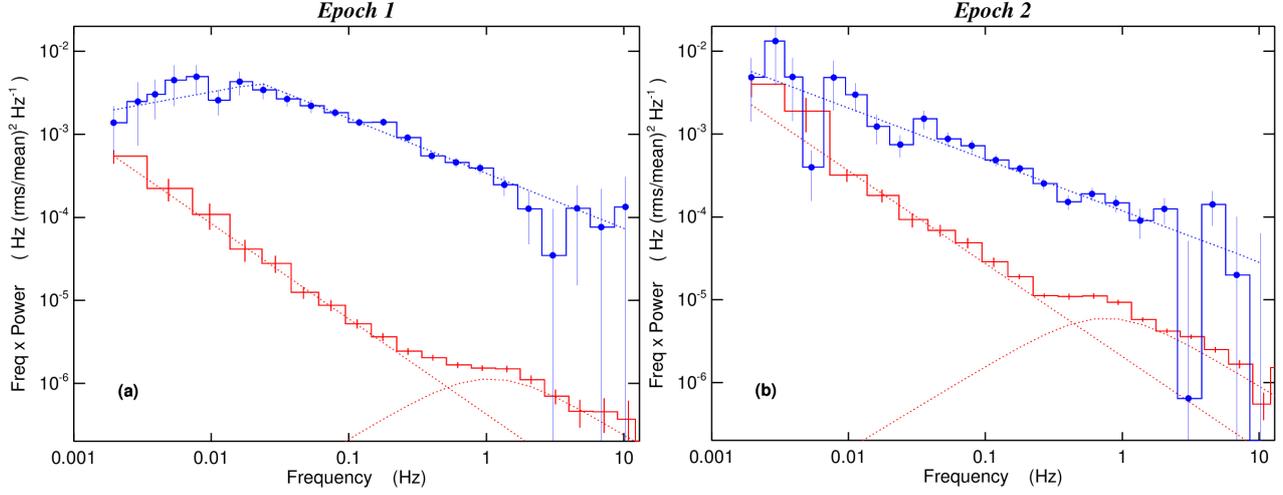

**Supplementary Figure 5: <u>Power spectral densities (PSDs) of the strictly simultaneous light curve epochs.</u>**

**The PSDs are plotted for Epoch 1 (a) and Epoch 2 (b) in $\nu P_\nu$ units with rms$^2$ normalisation and white noise contribution is removed. Optical and X-ray PSDs are plotted in red and blue, respectively. The dotted curves show the fitted model power law components to both optical and X-rays and zero-centred Lorentzian components in the case of the optical PSDs. Uncertainties denote 1 mean std. error for each bin.**

## Summary of known source parameters and related inferences

V404 Cygni is known to have a mass function of 6.08 $M_\odot$ and an orbital period of 6.47 days[16]. Its distance has been measured based upon radio parallax measurements to be 2.39 kpc[17]. The donor spectral type remains somewhat ambiguous, ranging between K0 and K3[18—20]. Combined with uncertain corrections for accretion disc contamination, the present ranges on the inclination and compact object mass are 56—67 deg and $M_{BH}$=8—12 $M_\odot$, respectively[21].

Assuming a jet bulk speed of β (relative to *c*), and an inclination angle *i*, the time delay τ between X-ray emission from close to the black hole and subsequent optical synchrotron emission from an elevation *d* along the jet is given by τ=*d*/β*c* × [1-βcos(*i*)]. Compact jet speeds are thought to be characterised by β values of around ~0.9[22]. For V404 Cygni, Tetarenko et al.[23] measure typically lower values β≲0.65, but note that brighter ejecta tend to show faster speeds. Another important caveat is that their measurements refer to discrete ejecta, with the underlying compact jet (of more relevance here) remaining unconstrained. In any case, with β constrained to be less than 1, our measurement of τ provides an upper limit to the elevation *d* of the optical jet base, modulo a correction factor due to the inclination angle dependent bracket term. For the range of *i* quoted above (i.e. assuming the jet axis is similarly inclined to the disc axis), this correction is of order unity. There is no evidence at present for the jet to be angled much closer to the line-of-sight.[23] Assuming $\beta \sim 0.5$, $\tau=0.1$ s corresponds to an elevation $d=700—800\ R_S$. If speeds associated with the compact jet are higher, $\beta \sim 0.9$ (for instance), this value is only boosted by a factor of ~2. On the other hand, accounting for non-zero acceleration of the plasma between the launch point and the dissipation zone is expected to decrease $d$ further. In summary, our estimates provide robust upper limits to the elevation of optical jet emission zone, to within correction factors of order unity.

We finally note the presence of a significant disc wind that has been observed in V404 Cygni[24,25]. Detailed analysis of the cold (optically-emitting) phase of this wind places its formation region at the outer accretion disc, which has an extent of ~$10^{12}$ cm. The minimum size scale of the wind launching zone is of order light-seconds[25]. These sizes are longer than the characteristic timescales associated with the sub-second flaring, arguing against reprocessing in this wind for their origin. It is worth noting that the bluer colour and the longer timescales associated with the *slower variations* may be compatible with an outer disc or wind reprocessing scenario, though a non-thermal origin is also plausible at present[5].

**Comparison to other sources with fast timing data**

The optical/X-ray timing correlations for V404 Cygni are different in several respects from those seen in other low mass X-ray binaries (LMXBs) in outburst, including XTE J1118+480[26], GX 339-4[9] and Swift J1753.5-0127[27]. In all these cases, a correlation between the light curves on timescales of ~0.1-10 s was identified and associated with either a compact jet or an inner hot flow. These sources also showed an optical vs. X-ray *anti-correlation* on characteristic times of ≲10 s, associated with sharing of accretion energy between various components, though there is much diversity even amongst these sources[28—32]. Data on other black hole binaries[33] is currently too insensitive to probe the fast correlation features in question here.

Swift J1753.5-0127 can be well explained by synchrotron self-Compton in a hot flow[31]. Consistent with this, the source is known to be radio-quiet with no strong requirement for a jet contributing much to the optical regime[27]. Thus, we do not consider it further in the context of the discussion below. In any case, we can rule out a synchrotron self Compton model where the optical photons seed Comptonisation, because in such a case, we would expect the optical to precede the X-rays, and the two bands to be anti-correlated. We do not observe either of these effects in V404 Cygni.

GX 339-4 is closest matched to what we see in V404 Cygni. A fast optical delay component was observed in GX 339-4 at a similar timescale of ~+0.1 $s^{9,32}$ and associated with the base of the optically-emitting jet at a few $10^3$ Schwarzschild radii from the black hole. Consistent with this, similar infrared lags have been found supporting an inner jet origin[22,34], and the detection of a broadband synchrotron spectral break has also provided an estimate of the average size of the optical emission zone in the jet to be ~0.1 light-seconds[35]. The overall optical variability of V404 Cygni, however, is dominated by the long timescale, slow variations. This is seen from the slow time lags peaking at timescales of order ~10 s (Supplementary Fig. 2, ref. 10), as well as the fact that the optical variability power spectra of V404 Cygni are much steeper[5,36,37] than GX 339-4 and other sources[38]. The reasons for this difference are unclear, but V404 Cygni stands out in terms of its long orbital period (6.5 days[16]; as compared to 1.7 days for GX 339-4[39]) and hence larger disc and/or extended corona, which plausibly implies longer variability timescales. In this sense, the source has several similarities[36] to the variations seen in the even longer (33-day) period system GRS 1915+105. The two other LMXBs referred to above (XTE J1118+480 and Swift J1753.5-0127) both have shorter orbital periods. This could explain the dominance of the longer variations in V404 Cygni.

With regard to the comparison with XTE J1118+480 – this source shows a continuum of optical time lags extending from ~10 ms to several seconds[40], with no strong preference for any one single time lag. XTE J1118+480 was observed during outburst decline with falling optical, infrared and X-ray emission. This behaviour may be similar to the time lags that we observe in V404 Cygni during Epoch 1, before the compact jet began to strengthen. In fact, Malzac et al.[40] found an optical time lag decreasing with Fourier frequency ($\nu$) as $\nu^{-0.8}$ for XTE J1118+480, while for Epoch 1, we find a dependence of $\nu^{-0.9}$ (Supplementary Fig. 2), although we note that the weak interband coherence above ~0.1 Hz prevents us from drawing any firm conclusions in this regard. The peak of the optical vs. X-ray coherence in V404 Cygni is shifted to lower Fourier frequency (longer timescales), presumably again a consequence of the long period noted above. But otherwise, this comparison makes a clear prediction that a ~+0.1 s lag could also be found in XTE J1118+480 at a different stage during outburst, when the optical jet emission was rising rather than declining. This can be tested during future outbursts.

Finally, we note that V404 Cygni is known to show significant multiwavelength flickering even in quiescence[41–45], but the observed variations in this case range from seconds to days, i.e. orders of magnitude longer timescales than under consideration here. Such variability has not been observed in GX 339-4, which could simply be a reflection of the fact that V404 Cygni is relatively bright and well studied (e.g. the source quiescent mags are $g'$=19.8 and $i'$=16.6[41] as compared $r$=20.1 for GX 339-4[46]). The origin of this variability remains to be understood but may be related to residual accretion and ejection of clumps within a geometrically-thick optically-thin accretion flow in between accretion disc buildup episodes[42,47].

1. Itoh, R. et al. A measurement of interstellar polarization and an estimation of Galactic extinction for the direction of the X-ray black hole binary V404 Cygni. *Publ. Astron. Soc. Jpn.* **69**, 25—33 (2017).
2. Casares, J., Charles, P. A., Naylor, T., & Pavlenko, E. P. Optical Studies of V:404-CYGNI the X-Ray Transient GS:2023+338 - Part Three - the Secondary Star and Accretion Disc. *Mon. Not. R. Astron. Soc.* **265**, 834—852 (1993).
3. Rahoui, F. et al. The nova-like nebular optical spectrum of V404 Cygni at the beginning of the 2015 outburst decay. *Mon. Not. R. Astron. Soc.* **465**, 4468—4481 (2017).
4. Cardelli, J. A., Clayton, G. C. & Mathis, J. S., The relationship between infrared, optical, and ultraviolet extinction. *Astrophys. J.* **345**, 245—256 (1989).
5. Gandhi P. et al. Furiously fast and red: sub-second optical flaring in V404 Cyg during the 2015 outburst peak. *Mon. Not. R. Astron. Soc.* **459**, 554—572 (2016).
6. Walton, D.J. et al. Living on a flare: Relativistic Reflection in V404 Cyg Observed by *NuSTAR* During its Summer 2015 Outburst. *Astrophys. J.* **839**, 110—132 (2017).
7. Motta, S. et al. The black hole binary V404 Cygni: an obscured AGN analogue. *Mon. Not. R. Astron. Soc.* **468**, 981—993 (2017).
8. Vaughan, B. A., & Nowak, M. A. X-Ray Variability Coherence: How to Compute It, What

It Means, and How It Constrains Models of GX 339-4 and Cygnus X-1. *Astrophys. J. Lett.* **474**, L43—L46 (1997).
9. Gandhi, P. et al. Rapid optical and X-ray timing observations of GX 339-4: flux correlations at the onset of a low/hard state. *Mon. Not. R. Astron. Soc.* **390**, L29—L33 (2008).
10. Gandhi, P. et al. Correlated Optical and X-ray variability in V404 Cyg. *Astron. Telegr.* **7727** (2015).
11. Belloni, T., & Hasinger, G. An Atlas of aperiodic variability in HMXB. *Astron. & Astrophys.* **230**, 103—119 (1990).
12. van der Klis, M. *Stat. Challenges in Modern Astron.* **II**. *Eds. Babu G.J. & Feigelson E.D.* 321 (1997).
13. Bachetti, M. et al. No Time for Dead Time: Timing Analysis of Bright Black Hole Binaries with NuSTAR. *Astrophys. J.* **800**, 109—120 (2015).
14. Malzac, J. The spectral energy distribution of compact jets powered by internal shocks. *Mon. Not. R. Astron. Soc.* **443**, 299—317 (2014).
15. Jenke, P.A. et al. Fermi GBM Observations of V404 Cyg During its 2015 Outburst. *Astrophys. J.* **826**, 37—47 (2016).
16. Casares, J. et al. A 6.5-day periodicity in the recurrent nova V404 Cygni implying the presence of a black hole. *Nature* **355**, 614—617 (1992).
17. Miller-Jones, J. et al. The first accurate parallax distance to a black hole. *Astrophy. J.* **706**, L230—L234 (2009).
18. Casares, J., Charles, P.A. Optical studies of V404 Cyg, the X-ray transient GS 2023+338. IV. The rotation speed of the companion star. *Mon. Not. R. Astron. Soc.* **271**, L5—L9 (1994).
19. Gonzalez Hernandez, J.I. et al. Chemical Abundances of the Secondary Star in the Black Hole X-Ray Binary V404 Cygni. *Astrophys. J.* **738**, 95—107 (2011).
20. Khargharia, J., Froning, C. S. & Robinson, E. L., Near-infrared Spectroscopy of Low-mass X-ray Binaries: Accretion Disk Contamination and Compact Object Mass Determination in V404 Cyg and Cen X-4. *Astrophys. J.* **716**, 1105—1117 (2010).
21. Casares J., Jonker, P.G. Mass measurements of stellar and intermediate-mass black holes. *Space Sci. Rev.* **183**, 223—252 (2014).
22. Casella, P. et al. Fast infrared variability from a relativistic jet in GX 339-4. *Mon. Not. R. Astron. Soc.* **404**, L21—L25 (2010).
23. Tetarenko, A.J. et al. Extreme Jet Ejections from the Black Hole X-ray Binary V404 Cygni. *Mon Not. R. Astron. Soc.* **469,** 3141—3162 (2017).
24. King, A.L. et al. High resolution Chandra HETG spectroscopy of V404 Cygni in outburst. *Astrophys. J.* **813**, L37—L44 (2015).
25. Muñoz-Darias, T. et al. Regulation of black-hole accretion by a disk wind during a violent outburst of V404 Cygni. *Nature*. **534**, 75—78 (2016).
26. Kanbach, G. et al. Correlated fast X-ray and optical variability in the black-hole candidate XTE J1118+480. *Nature* **414**, 180—182 (2001).
27. Durant, M. et al. Multiwavelength spectral and high time resolution observations of SWIFT J1753.5-0127: new activity? *Mon. Not. R. Astron. Soc.* **392**, 309—324 (2009).
28. Malzac, J. et al. Jet-disc coupling through a common energy reservoir in the black hole XTE J1118+480. *Mon. Not. R. Astron. Soc.* **351**, 253—264 (2004).
29. Durant, M. et al. High time resolution optical/X-ray cross-correlations for X-ray binaries: anticorrelation and rapid variability. *Mon. Not. R. Astron. Soc.* **410**, 2329—2338 (2011).

30. Hynes, R. I., et al. The remarkable rapid X-ray, ultraviolet, optical and infrared variability in the black hole XTE J1118+480. *Mon. Not. R. Astron. Soc.* **345**, 292—310 (2003).
31. Veledina, A., Poutanen, J., & Vurm, I. A Synchrotron Self-Compton-Disk Reprocessing Model for Optical/X-Ray Correlation in Black Hole X-Ray Binaries. *Astrophys. J. Lett.* **737**, L17—L21 (2011).
32. Gandhi, P. et al. Rapid optical and X-ray timing observations of GX339-4: multicomponent optical variability in the low/hard state. *Mon. Not. R. Astron. Soc.* **407**, 2166—2192 (2010).
33. Pahari, M., Gandhi, P., Charles, P. A., Kotze, M. M., Altamirano, D., & Misra, R. Simultaneous Optical/X-ray study of GS 1354-64 (=BW Cir) during hard outburst: evidence for optical cyclo-synchrotron emission from the hot accretion flow. *Mon. Not. R. Astron. Soc.* **469**, 193—205 (2017).
34. Kalamkar, M. et al. Detection of the first infrared quasi-periodic oscillation in a black hole X-ray binary. *Mon. Not. R. Astron. Soc.* **460**, 3284—3291 (2016).
35. Gandhi, P. et al. A Variable Mid-Infrared Synchrotron Break associated with the Compact Jet in GX 339—4. *Astrophys. J.* **740**, L13—L19 (2011).
36. Kimura, M. et al. Repetitive patterns in rapid optical variations in the nearby black-hole binary V404 Cygni. *Nature* **529**, 54—58 (2016).
37. Hynes, R. I., Robinson, E. L. & Morales, J. Rapid Optical Photometry of V404 Cyg. *Astron. Telegr.* **#7677** (2015).
38. Gandhi, P. The Flux-Dependent rms Variability of X-Ray Binaries in the Optical. *Astrophys. J. Lett.* **697**, L167—L172 (2009).
39. Hynes, R. I., et al. Dynamical Evidence for a Black Hole in GX 339-4. *Astrophys. J. Lett.* **583**, L95—L98 (2003).
40. Malzac, J., Belloni, T., Spruit, H. C., & Kanbach, G., The optical and X-ray flickering of XTE J1118+480. *Astron. & Astrophys.* **407**, 335—345 (2003).
41. Shahbaz, T. et al. Multicolour observations of V404 Cyg with ULTRACAM. *Mon. Not. R. Astron. Soc.* **346**, 1116—1124 (2003).
42. Zurita, C., Casares, J., Shahbaz, T. Evidence for Optical Flares in Quiescent Soft X-Ray Transients. *Astrophys. J.* **582**, 369—381 (2003).
43. Bernardini, F. & Cackett, E. M. Characterizing the quiescent X-ray variability of the black hole low-mass X-ray binary V404 Cyg. *Mon. Not. R. Astron. Soc.* **439**, 2771—2780 (2014).
44. Rana, V. et al. Characterizing X-Ray and Radio Emission in the Black Hole X-Ray Binary V404 Cygni during Quiescence. *Astrophys. J.* **821**, 103—112 (2016).
45. Plotkin, R. M. et al. The 2015 Decay of the Black Hole X-Ray Binary V404 Cygni: Robust Disk-jet Coupling and a Sharp Transition into Quiescence. *Astrophys. J.* **834**, 104—122 (2017).
46. Shahbaz, T., Fender, R. & Charles, P.A. VLT optical observations of V821 Ara(=GX339-4) in an extended "off" state. *Astron. Astrophys.* **376**, L17—L21 (2001).
47. Shahbaz T. et al. ULTRACAM observations of the black hole X-ray transient XTE J1118+480 in quiescence. *Mon. Not. R. Astron. Soc.* **362**, 975—982 (2005).